\documentstyle[12pt]{article}

\def\be{\begin{equation}}
\def\ee{\end{equation}}

\begin{document}
\vspace*{.3cm}
\begin{flushright}
\large{CINVESTAV-FIS-01/96}
\end{flushright}
\begin{center}
\LARGE{\bf Isospin corrections to charmless semileptonic $B \rightarrow 
V$ transitions}
 \end{center}
\vspace{.8cm}
\begin{center}
\Large J. L. D\'\i az-Cruz$^1$, G. L\'opez Castro$^2$ and J. H. 
Mu\~noz$^{2,3}$\\ %\medskip
\vspace*{.4cm}
{\normalsize{\it $^1$ Instituto de F\'\i sica, Universidad Aut\'onoma de 
Puebla \\ 
 \vspace{-.1cm}\it Ap. Postal J-48, 72500 Puebla, Pue. MEXICO. \\
\it $^2$ Departamento de F\'\i sica, Cinvestav del IPN, Apdo. \\ 
\vspace{-.1cm}\it  Postal 14-740, 07000 M\'exico, D.F. MEXICO. \\
\it $^3$ Departamento de F\'\i sica, Universidad del Tolima, \\
\vspace{-.5cm} \it A. A. 546, Ibagu\'e. COLOMBIA}}
\vspace*{.4cm}
\end{center}

\thispagestyle{empty}
\centerline{ \bf Abstract}
\vspace{.3cm}
  We compute isospin corrections to the charmless semileptonic 
$B \rightarrow V$ transitions arising from $\rho-\omega$ mixing 
and discuss its relevance in the determination of $V_{ub}$.

 \vspace{.5cm} PACS numbers:  13.20.He, 12.15.Hh, 12.40.Vv. 11.30.Hv

\newpage
\setcounter{page}{1}
\vspace{2cm}

  The first measurement of an exclusive charmless semileptonic B decay  
has been reported recently by the CLEO Collaboration [1,4]. The 
yields for $B^+ \rightarrow \pi^0 l^+ \nu_l$ reported in Refs. [1,4] 
turn out to be strongly dependent on the theoretical models [2,3] used 
for the detection efficiencies, as it does the extraction of the CKM 
parameter $|V_{ub}|$ from these results.

   The $|V_{ub}|$ parameter can be determined from exclusive charmless 
semileptonic B 
decays and also from the end-point region of the lepton spectrum in 
inclusive semileptonic B decays. Althought one does not expect that 
exclusive 
measurements will provide a better determination of $|V_{ub}|$ than 
the inclusive measurements does [4], the former are important for several 
reasons. Besides the complementary determination of $|V_{ub}|$ provided by 
exclusive B decays, these measurements will allow to test available  
theoretical models for form 
factors (relativistic [2] and non-relativistic [3] quark models, QCD sum 
rules [5]).
   In this paper we are concerned with important isospin breaking 
corrections to $B \rightarrow V l \nu\ (V=\rho^+,\ \rho^0\ {\rm and}\  
\omega)$ decays which have not been considered in previous analysis.
 
There are several motivations to consider all sources of theoretical 
corrections in the calculation of charmless $B \rightarrow V$ 
semileptonic decays. From the experimental point of view,
 the study of these decays can be sensible 
improved at the B factory. In fact, actual measurements by the CLEO 
collaboration [6] provide the upper limit $B(B^- \rightarrow \rho^0 l^- 
\bar{\nu}_l) < (1.6 - 2.7) \times 10^{-4}$, which lies at the verge of 
theoretical predictions.

  Among the theoretical motivations we can mention the following: $(a)$ the 
decay rate of $B\rightarrow \rho l \nu$ provides 3 -- 14 \% 
[2,3,5] of the inclusive $B \rightarrow X_u l \nu_l$ decays, $(b)$ the 
ratio of decay widths 
for  $(B \rightarrow V l \nu)/(B \rightarrow \pi l \nu)$ and the 
polarization of the daugther vector mesons can be used to discriminate 
between form factor models and, $(c)$ recent proposals [7] 
indicate that we could achieve a model-independent determination of 
$|V_{ub}|$ from $B\rightarrow \rho l \nu$, at the level of 10\%, by 
using HQET techniques and SU(3) flavor symmetry.

  Let us now discuss the relevance of isospin breaking   
corrections to semileptonic $B$ decays. The experimental measurements 
reported by the CLEO 
collaboration on $B \rightarrow \pi l \nu$ [1,4] and the upper limit set 
on $B \rightarrow \rho l \nu$ [6] rely on the following assumptions:
\begin{eqnarray}
\Gamma(\bar{B}^0 \rightarrow \rho^+ l^- \bar{\nu}) & = & 2 \Gamma(B^- 
\rightarrow \rho^0 l^- \bar{\nu}) \nonumber \\
& \approx & 2 \Gamma(B^- \rightarrow \omega l^- \bar{\nu}) 
\end{eqnarray}
and 
\begin{equation}
\Gamma(\bar{B}^0 \rightarrow \pi^+ l^- \bar{\nu}) = 2\Gamma(B^- 
\rightarrow \pi^0 l^- \bar{\nu}) 
\end{equation}
 
  The first row in Eqs. (1) and (2) are valid in the limit of exact isospin 
symmetry, while the second equality in Eq. (1) follows from the 
$u\bar{u}$ content of $\rho^0$ and $\omega$ mesons in the limit of exact 
isospin symmetry. Corrections to these relations arise from electromagnetic 
radiative corrections and from the $u-d$ quark mass difference.
   Electromagnetic radiative corrections [8] to $B \rightarrow \rho l 
\nu$ and phase space corrections due to the physical masses of the 
mesons are expected to be negligible 
due to the large B-meson mass.  Isospin corrections to Eq. (1) induced by 
$\rho-\omega$ mixing (which arise from the $u-d$ quark mass difference) are 
the subject of this paper. This correction turns out to be large due to 
the small difference in the vector meson masses ($m_{\omega} - m_{\rho}  
\approx 12\ {\rm MeV}$) and the large difference in their decay widths 
($\Gamma_{\omega} <<\Gamma_{\rho}$)\footnote{ Isospin corrections to Eq. 
(2), arising from $\pi^0-\eta$ mixing, are negligible [9].}.

Let us proceed with our calculation. In the limit of isospin symmetry, 
$\rho^0$ and $\omega$ are isospin eigenstates with flavor content: 
$\rho^I=(u\bar{u}-d\bar{d})/\sqrt{2}$ and
$\omega^I=(u\bar{u}+d\bar{d})/\sqrt{2}$. Since the spectator quark in 
$B^- \rightarrow (\rho^0, \ \omega) l^- \bar{\nu}_l$\ is $\bar{u}$, the 
$\rho^0$ and $\omega$ mesons are produced from their $u\bar{u}$ quark 
content. This provides the following equality between hadronic matrix 
elements:
\begin{equation}
<\rho^I | \bar{u}\gamma_{\mu}(1-\gamma_5)b | B^->\ =
\ <\omega^I | \bar{u}\gamma_{\mu}(1-\gamma_5)b | B^->
\end{equation}
which leads to the second row in Eq. (1).

   When we introduce isospin breaking, $\rho^I$ and $\omega^I$ get mixed 
into physical states $\rho^0$ and $\omega$, namely:
\begin{eqnarray}
 \omega  &=& \omega^I - \epsilon' \rho^I \nonumber \\   
   & = & \frac{1}{\sqrt{2}}(1-\epsilon')u\bar{u} 
+\frac{1}{\sqrt{2}}(1+\epsilon')d\bar{d} \\
 \rho^0  &=& \rho^I + \epsilon\ \omega^I \nonumber \\   
   & = & \frac{1}{\sqrt{2}}(1+\epsilon)u\bar{u} 
+\frac{1}{\sqrt{2}}(-1+\epsilon)d\bar{d} 
\end{eqnarray}
where  $\epsilon$($\epsilon'$) are the contributions 
of $\rho-\omega$ mixing, with the following expressions: 
\begin{eqnarray}
\epsilon &=& \frac{m_{\rho\omega}^2}{m_{\rho}^2-m_{\omega}^2 + 
im_{\omega}\Gamma_{\omega}} \\   
\epsilon' &=& \frac{m_{\rho\omega}^2}{m_{\omega}^2-m_{\rho}^2 + 
im_{\rho}\Gamma_{\rho} }
\end{eqnarray}
where $m_{\rho\omega}^2 = (-3.67 \pm 0.30) \times 10^{-3}\ {\rm GeV}^2$ 
[10] is the strenght of $\rho-\omega$ mixing.

   After including $\rho-\omega$ mixing, Eq. (3) becomes:
\begin{equation}
<\rho^0 | \bar{u}\gamma_{\mu}(1-\gamma_5)b | B^->\ =\  
\frac{1+\epsilon}{1-\epsilon'}
<\omega | \bar{u}\gamma_{\mu}(1-\gamma_5)b | B^->.
\end{equation}

   From this equation we can obtain the ratio for $\rho^0$ and $\omega$ 
production in $B^-$ decays. If we use $m_{\rho} = 757.5$ MeV and 
$\Gamma_{\rho} =142.5$ MeV as obtained from a recent fit to the pion 
form factor [10] and the $\omega(782)$ parameters from [11], we get: 
\begin{equation}
\frac{\Gamma (B^- \rightarrow \rho^0 l^- \bar{\nu})}
{\Gamma (B^- \rightarrow \omega l^- \bar{\nu})} = \left| 
\frac{1+\epsilon}{1-\epsilon'} \right|^2 \approx 1.172
\end{equation}
whereas the use of PDG values [11] for the $\rho^0$ parameters gives:
\begin{equation}
\frac{\Gamma (B^- \rightarrow \rho^0 l^- \bar{\nu})}
{\Gamma (B^- \rightarrow \omega l^- \bar{\nu})}  \approx 1.367
\end{equation}
which looks a rather large correction.

  Note that the ratio $2 \Gamma(B^- \rightarrow \rho^0 l^- \bar{\nu}_l)/ 
\Gamma(\bar{B}^0 \rightarrow \rho^+ l^- \bar{\nu}_l)$ is modified by 
almost the 
same amount as Eq. (9) does, since $\rho-\omega$ mixing affects only the 
neutral vector 
mesons. We would like to stress that $B \rightarrow (\rho,\omega) l \nu$ 
are affected by this correction 
regardless of the model used to describe the form factors of the $B 
\rightarrow 
V$ transition. Let us comment that the calculation of the corresponding form 
factors in refs. [2] and [3] assume explicitly $m_{u}=m_{d}$. Note that 
the ratio $\Gamma(B^- \rightarrow \rho^0 l^- \bar{\nu})/\Gamma(\bar{B}^0 
\rightarrow \rho^+ l^- \bar{\nu})$ plays for $B$ decays the same role as 
$\Gamma(K^+ \rightarrow \pi^0 l^+ \nu)/\Gamma(K_L \rightarrow \pi^- l^+ 
\nu)$ [9] does for $K$ decays  in order to test the flavor-symmetry 
breaking  corrections to form factors at zero momentum transfer (which 
are essential for the determination of $V_{us}$).

 The individual decay rates for $B^-$ semileptonic decays are afected by 
$\rho-\omega$ mixing as follows:
\begin{eqnarray}
\Gamma(B^- \rightarrow \rho^0 l^- \bar{\nu}_l) &=& |1+\epsilon|^2 
\Gamma^0( B^- \rightarrow \rho^0 l^- \bar{\nu}_l) \\
\Gamma(B^- \rightarrow \omega l^- \bar{\nu}_l) &=& |1-\epsilon'|^2 
\Gamma^0( B^- \rightarrow \omega l^- \bar{\nu}_l) 
\end{eqnarray}
where $\Gamma^0$ denote the decay rate without $\rho-\omega$ mixing. 
Since $|1+\epsilon| \approx 1.095$ and $|1-\epsilon'| \approx 1.011$ ( 
1.18 and 1.005 if the $\rho^0$ parameters of [11] are used, 
respectively), the values of $|V_{ub}|$ as extracted from $B^- 
\rightarrow \rho^0 l^- \bar{\nu}$ and $B\rightarrow \omega l^- \bar{\nu}$ 
would be, respectively, 10\% and 1\% (18\% and 0.5\%) higher if 
$\rho-\omega$ mixing were not included.

  It is straightforward to extend this analysis to strangeness-conserving 
$D^+ \rightarrow (\rho^0, \omega)$ semileptonic transitions. Since the 
$D^+ \rightarrow (\rho^0, \omega) l^+ \nu$ decays proceed through the 
elementary transition $c\rightarrow d l^+ \nu$, the spectator $\bar{d}$ 
quark get combined with the daugther $d$ quark to produce the $\rho^0$ 
and $\omega$ mesons. Using the $\rho^0$ parameters from Ref. [10] we get, 
after including $\rho-\omega$ mixing, the following ratio:
\begin{eqnarray}
\frac{\Gamma(D^+ \rightarrow \rho^0 l^+ \nu)}{\Gamma(D^+ \rightarrow \omega 
l^+ \nu)} & = & \left| \frac{-1+\epsilon}{1+\epsilon'} \right|^2 \nonumber \\
&\approx & 0.837\ (0.692)
\end{eqnarray}
{\em i.e.}, the correction due to $\rho-\omega$ mixing is similar as in 
$B$ decays but it goes in the opposite direction (the number in brackets 
is obtained for the $\rho,\ \omega$ parameters of Ref. [11]). The 
corresponding experimental information [11]  available for semileptonic $D$ 
decays is not precise enough to allow a test of Eq. (13).

   Finally, in order to trust our calculations, we can compute the ratio of 
decay widths for radiative decays of $\rho$ mesons, namely: $R\equiv 
\Gamma(\rho^0 \rightarrow \pi^0\gamma)/
\Gamma(\rho^+ \rightarrow \pi^+\gamma)$. In this case, the $\rho^0$ decay 
receives an additional contribution from $\rho-\omega$ mixing ($\rho^0 
\rightarrow \omega \rightarrow \pi^0\gamma$). The ratio $R$ is modified 
to become [10]:
\begin{equation}
R=\left| 1 + \frac{f_{\omega}}{f_{\rho}}\cdot \epsilon \right|^2 \left| 
\frac{\vec{k}_{\pi^0}}{\vec{k}_{\pi^+}} \right|^3
\end{equation}
where $\vec{k}_{\pi}$ is the pion momentum in the $\rho$ rest frame, and 
$em_{V}^2/f_V$ defines the vector-meson--photon coupling.

   Using $f_{\rho}=5.0$ and $f_{\omega}=17.0$ from $(\rho^0,\ \omega) 
\rightarrow e^+e^-$ decays, we obtain [10] 
\begin{equation}
R=1.77\ (2.40)
\end{equation}
where the number in brackets correspond to the $\rho^0$ parameters of Ref. 
[11]. The above result is in good agreement with the experimental 
value $R^{exp} = 1.78 \pm 0.49$ [11].

   In conclusion, the $\rho-\omega$ mixing induces a sizable correction 
to the isospin symmetry relations given in Eq. (1). This overall 
correction to $B \rightarrow (\rho, \omega) l^- \bar{\nu}$ is present 
regardless of the specific form factor model [2,3,5] used to describe the 
hadronic weak transition. The values of $V_{ub}$ as extracted from $B^- 
\rightarrow (\rho^0, \omega)l^- \bar{\nu}$ would be overestimated by 10 
\% and 1\% if $\rho-\omega$ mixing is not included.

\

{\bf Acknowledgements}

   One of the authors (JHM) acknowleges the financial support from 
COLCIENCIAS.

\

\end{document}